\documentclass[12pt]{article}
\usepackage{a4wide}
\usepackage{amssymb}
\usepackage{graphicx}
\begin{document}
{\renewcommand{\thefootnote}{\fnsymbol{footnote}}
\begin{center}
{\LARGE Tunneling dynamics in cosmological bounce models}\\
\vspace{1.5em}
Martin Bojowald\footnote{e-mail address: {\tt bojowald@psu.edu}}
and Brenda Jones\footnote{e-mail address: {\tt  brenda.jones@maine.edu}, new
  address: Department of Physics and 
    Astronomy, The University of Maine, 5709 Bennett Hall, Orono, ME 04469, USA}
\\
\vspace{0.5em}
Department of Physics,\\
The Pennsylvania State
University,\\
104 Davey Lab, University Park, PA 16802, USA\\
\vspace{1.5em}
\end{center}
}

\setcounter{footnote}{0}

\begin{abstract}
  Quasiclassical methods are used to define dynamical tunneling times in
  models of quantum cosmological bounces. These methods provide relevant new
  information compared with the traditional treatment of quantum tunneling by
  means of tunneling probabilities. As shown here, the quantum dynamics in
  bounce models is not secure from reaching zero scale factor, re-opening the
  question of how the classical singularity may be avoided. Moreover, in the
  examples studied here, tunneling times remain small even for large barriers,
  highlighting the quantum instability of underlying bounce models.
\end{abstract}

\section{Introduction}

A possible scenario of cosmological dynamics that avoids the classical
big-bang singularity is given by bounces in modified or quantized models of
gravity. If matter or gravity behave non-classically at high curvature, their
features may present a barrier to evolution such that unbounded curvature or
zero scale factor of an isotropic model are not reached. Instead, the scale
factor may approach a non-zero constant value asymptotically, or it may bounce
back to macroscopic values after a moment of maximal collapse \cite{BounceReview}.

Most models of this kind are evaluated in a way that is essentially classical
(or, in more precise terms, leading-order quasiclassical), modifying the
Friedmann equation without directly representing it through quantum
operators. While it is technically possible to evolve a quantum state within a
spatially homogeneous model of quantum cosmology, the uncertain nature of
quantum space-time physics makes it hard to extend such results to
inhomogeneous perturbations, such as the curvature or density distribution
relevant for determining the primordial state underlying the cosmic microwave
background. Moreover, we hardly have any information about possible
restrictions on the quantum state the universe may be in at early times. It is
therefore unclear how significant properties of specific evolved wave
functions can be, beyond what the leading-order quasiclassical dynamics of the
scale factor in a modified Friedmann equation would provide.

A key property in which quantum dynamics differs from classical physics is
given by the tunneling effect. In early-universe cosmology, this effect
suggests that barriers to evolution as they may be used in bounce models are
not absolute. Finite barriers may slow down evolution across the barrier, but
they do not completely prevent it. A direct classical treatment of a simple
modified Friedmann equation ignores the tunneling effect. It could therefore
be possible that quantum evolution remains singular even if a first-order
quasiclassical treatment might suggest a bounce. This question has been raised
and analyzed in \cite{OscillatingInstability}, in particular for models of
loop quantum cosmology, using the method of Euclidean saddle points in order
to estimate tunneling probabilities. 

A direct quantum treatment by wave functions is available in quantum cosmology
and would be sensitive to tunneling effects. However, the usual stationary
setting familiar from quantum mechanics might miss crucial features of the
essential dynamical nature of the early universe. Moreover, the collapse of a
single universe is conceptually different from a scattering process of an
extended stream of particles, as assumed in standard tunneling
calculations. Given the unknown nature of the quantum state of the universe,
reliable conclusions in quantum cosmology require a sufficiently large set of
states, parameterized in a suitable way that does not assume specific physical
properties.

Here, we introduce methods of non-adiabatic quantum dynamics to bounce models
in early-universe cosmology. These methods have appeared independently in
various fields
\cite{VariationalEffAc,GaussianDyn,EnvQuantumChaos,QHDTunneling,CQC,CQCFieldsHom}
and have recently been shown to capture details tunneling effects in extended
quasiclassical formulations \cite{Bosonize,EffPotRealize,Ionization}. In
particular, these methods reveal the entire process of tunneling and can be
used to determine characteristic parameters such as tunneling times. (See for
instance \cite{TunnelingReview,BarrierTime,AttoclockReview} for discussions of
the concept of tunneling times.) In their present form, they rely on
semiclassical approximations, but based on a systematic expansion that can be
applied to higher than the leading orders often assumed in heuristic
discussions of bounce models.

\section{Models}

Our analysis is based on an anharmonic reformulation of the Friedmann dynamics
derived in \cite{Harmonic,NonBouncing}. Starting with
\begin{equation}
  \left(\frac{\dot{a}}{a}\right)^2= \frac{8\pi G}{3}\rho
\end{equation}
for vanishing spatial curvature, we first replace $\dot{a}$ with its canonical
momentum, $p_a=-(3/4\pi G) V_0 a\dot{a}$ where $V_0$ is the coordinate volume
of a finite region in space. Specifying the matter density $\rho$ as given
by a free, massless scalar field $\phi$ with momentum $p_{\phi}$,
\begin{equation}
  \rho=\frac{1}{2}\frac{p_{\phi}^2}{a^6}\,,
\end{equation}
we then note that the Friedmann equation implies the expression
\begin{equation} \label{pphia}
  p_{\phi}(a,p_a)=\pm \sqrt{\frac{3}{4\pi G}} |ap_a|
\end{equation}
for $p_{\phi}$ quadratic in the canonical variables $a$ and $p_a$.

\subsection{Harmonic systems}

Hamilton's equations generated by $p_{\phi}(a,p_a)$ determine how $a$ and
$p_a$ evolve with respect to the scalar value $\phi$ as a formal time
parameter. The quadratic nature of (\ref{pphia}) shows that this evolution is
essentially harmonic, except for the absolute value taken for $ap_a$ which
might suggest that the Hamiltonian is not strictly quadratic. However, since
the sign of $ap_a$ is preserved by evolution generated by a Hamiltonian
proportional to $ap_a$, $p_{\phi}$ may be replaced with a Hamiltonian of the
form (\ref{pphia}) in which the absolute value has been dropped. Upon
quantization, one should then impose the condition that a given initial state
is supported either on the positive or on the negative part of the spectrum of
$ap_a$.

More generally, the quadratic nature is preserved by any canonical
transformation of $(a,p_a)$ in which the new configuration variable is related
to $a$ by a power law. A convenient definition that absorbs some of the
constant parameters introduces a class of new canonical variables $Q$ and $P$
by
\begin{equation} \label{QP}
 |Q|=\frac{3(\ell_0a)^{2(1-x)}}{8\pi G(1-x)} \quad,\quad
 P=-\ell_0^{1+2x}a^{2x}\dot{a}
\end{equation}
where $\ell_0=V_0^{1/3}$ and $x$ is a free parameter that determines a
particular canonical transformation. The absolute value of $Q$ indicates that
we may conveniently extend this variable to the full real range, including
negative values, if we combine the dependence on the scale factor with a sign
factor that determines the orientation of space. This choice simplifies
quantization because it removes a boundary from the classical phase space.

Also transforming $\phi$ to
\begin{equation} \label{lambda}
 \lambda= \sqrt{16\pi G/3}\:|1-x| \phi
\end{equation}
for a given $x$, we obtain the simple expression
\begin{equation} \label{H}
 H=p_{\lambda}=\pm|QP|
\end{equation}
for the momentum of $\lambda$, identified with the Hamiltonian $H$ for
evolution of $Q$ and $P$ with respect to $\lambda$.

So far, we have reformulated classical dynamics of a spatially flat, isotropic
universe with a specific matter ingredient. A possible modification of this
dynamics has been suggested by loop quantum cosmology \cite{LivRev,ROPP}, in
which the $\lambda$-Hamiltonian is no longer quadratic in canonical
variables but instead given by
\begin{equation}
 H_{\delta}=\pm\frac{|Q\sin(\delta P)|}{\delta}
\end{equation}
where $\delta$ characterizes the magnitude of quantum effects. While this
Hamiltonian is non-polynomial in canonical variables, it can be mapped to a
linear expression if non-canonical basic variables are used that form a
suitable Lie algebra. Many of the formal properties that characterize a
harmonic system are then preserved.

Specifically, in addition to $Q$, we use the complex expression
\begin{equation} \label{J}
  J=Q\exp(i\delta P)
\end{equation}
instead of the momentum $P$. The real and imaginary parts of $J$, together
with $Q$, provide three real functions that, based on their mutual Poisson
brackets
\begin{equation} \label{QJ}
 \{Q,{\rm Re}J\}= -\delta {\rm Im}J\quad,\quad \{Q,{\rm Im}J\}=\delta{\rm
   Re}J\quad,\quad \{{\rm Re}J,{\rm Im}J\}=\delta Q\,,
\end{equation}
can be identified with the generators of the Lie algebra
${\rm sl}(2,{\mathbb R})$ \cite{BouncePert}. (See also
\cite{GroupLQC,CVHComplexifier,CVHPolymer,CVHProtected,CoarseGrainSU11,RenormSU11}
for appearances of this algebra or related versions in quantum cosmology.) In
these variables, the Hamiltonian
\begin{equation}\label{Hdelta}
 H_{\delta}=\pm\frac{|{\rm Im}J|}{\delta}
\end{equation}
is linear (again, up to the absolute value).

\subsection{Analog oscillators}

While the systems derived so far are harmonic in a formal sense, their Hamiltonians do not
appear in standard form with a kinetic energy and a quadratic potential. It is
easy to see that the
Hamiltonian $H$ can be mapped to an inverted harmonic oscillator by a linear
canonical transformation of $Q$ and $P$. The family $H_{\delta}$, however, is
harder to bring to standard form in this way because it is not based on
canonical variables.

We therefore apply a different procedure to bring the Hamiltonians closer to
standard form, following \cite{NonBouncing}. Starting with the equations of
motion they generate, we will determine suitable energy expressions together
with additional constraints. For $H$, the canonical equations of motion with
respect to $\lambda$ are given by
\begin{equation} \label{QdotPdot}
  \dot{Q}=Q \quad \mbox{and}\quad \dot{P}=-P\,.
\end{equation}
The first equation directly implies the second-order equation of motion
$\ddot{Q}=\dot{Q}=Q$ which can also be obtained from the analog Hamiltonian
\begin{equation} \label{Hanalog}
 H_{\rm analog}=\frac{1}{2}(\pi^2-Q^2)\,.
\end{equation}
The momentum $\pi$ is distinct from the original momentum $P$, and it is
constrained by the fact that $H_{\rm analog}$ implies $\pi=\dot{Q}$ while we
also have $\dot{Q}=Q$ according to the original equations of motion. We can
therefore express the original dynamics for $Q$ by the standard inverted harmonic
oscillator (\ref{Hanalog}), subject to the constraint $H_{\rm analog}=0$
imposed on initial values.

The Hamiltonian $H_{\delta}$ generates equations of motion for the algebra
generators $Q$, ${\rm Re}J$ and ${\rm Im}J$:
\begin{equation} \label{Qdot}
  \dot{Q}={\rm Re}J
\end{equation}
and
\begin{equation} \label{Jdot}
  \frac{{\rm d}{\rm Re}J}{{\rm d}\lambda}=Q\,.
\end{equation}
The third equation of motion simply states that ${\rm Im}J$ is
conserved. While (\ref{QdotPdot}) is modified, we have the same second-order
equation as before, $\ddot{Q}=Q$. Therefore, we may use the same analog
Hamiltonian (\ref{Hanalog}). However, since $\dot{Q}\not=Q$ now, there is no
constraint that determines the value of $H_{\rm analog}$. There is now an
independent momentum $\pi$ canonically conjugate to $Q$ and not strictly
related to it.

The brackets (\ref{QJ}) show that we may use the equation
\begin{equation}
  \pi=\frac{{\rm Re}J}{\delta^2H_{\delta}}
\end{equation}
as an analog momentum, noting that ${\rm Im}J=-\delta H_{\delta}$ (up to a sign
choice) is conserved for evolution generated by $H_{\delta}$. Therefore,
$\pi=\delta^{-2}\dot{Q}/H_{\delta}$. If we define the analog Hamiltonian
\begin{equation}
  H_{\rm analog}=\frac{1}{2}  \left(\delta^2H_{\delta}
    \pi^2-\frac{1}{\delta^2H_{\delta}}Q^2\right)\,, 
\end{equation}
its Hamilton equations are equivalent to the equations of motion (\ref{Qdot})
and (\ref{Jdot}). A canonical transformation from $(Q,\pi)$ to
\begin{equation}
  \tilde{Q}=\frac{Q}{\delta\sqrt{H_{\delta}}} \quad,\quad
  \tilde{\pi}=\delta\sqrt{H_{\delta}} \pi
\end{equation}
(assuming, without loss of generality, $H_{\delta}>0$) brings this Hamiltonian
to the standard form of the inverted harmonic oscillator.

The value of the momentum $\pi$ is not strictly determined by $Q$, but it is
related to the Hamiltonian by a reality condition, based on the definition
(\ref{J}) of $J$ as a complex expression. This definition implies that
\begin{equation}
  |J|^2=({\rm Re}J)^2+({\rm Im}J)^2=Q^2\,.
\end{equation}
Therefore,
\begin{equation}
 \delta^2 \pi^2=\frac{({\rm Re}J)^2}{({\rm Im}J)^2}=\frac{Q^2}{({\rm
     Im}J)^2}-1 =\frac{Q^2}{\delta^2H_{\delta}^2}-1
\end{equation}
or
\begin{equation}
  \tilde{\pi}^2= \tilde{Q}^2-H_{\delta}\,.
\end{equation}
We obtain the value of the analog Hamiltonian given by
\begin{equation} \label{Hvalue}
  H_{\rm
    analog}=\frac{1}{2}\left(\tilde{\pi}^2-\tilde{Q}^2\right)=-\frac{1}{2}H_{\delta}=
  -\frac{1}{2}p_{\lambda}\,.
\end{equation}
Since $p_{\lambda}$ is a free parameter, there is no sharp constraint on
$H_{\rm analog}$ in this model.

For small $Q$, the behavior of the harmonic systems as models of quantum
cosmology may be modified, depending on quantization ambiguities
\cite{NonBouncing}. In particular, spatial discreteness may imply that the
linear expression for $\ddot{Q}$ is replaced by $\ddot{Q}=Q^2{\rm sgn}Q$ once
$Q$ is sufficiently small. The corresponding analog Hamiltonian has a cubic
potential,
\begin{equation}
  H_{\rm cubic}= \frac{1}{2} \tilde{\pi}^2-\kappa |\tilde{Q}|^3
\end{equation}
with some constant $\kappa$.  Since we are interested in quantum tunneling at
small $Q$, our main tunneling analysis will be based on this Hamiltonian. In
addition, because the power-law form may be subject to quantization choices,
we will consider the examples of a quartic potential and an anharmonic
potential with quadratic and quartic contributions. From now on, we will
suppress tildes in the variables of $H_{\delta}$ to simplify the notation.

\subsection{Quasiclassical extensions}

The initial Hamiltonian $H=QP$ can easily be quantized to
$\hat{H}=\frac{1}{2}(\hat{Q}\hat{P}+\hat{P}\hat{Q})$, choosing a standard symmetric
ordering. Quasiclassical formulations based on canonical effective equations
consider evolution generated by an effective Hamiltonian $H_{\rm
  eff}=\langle\hat{H}\rangle$ in a suitable class of states used to compute
the expectation value.

For a quadratic Hamilton operator, the effective
Hamiltonian can be derived for any state, provided the latter is given not by
a wave function but by a collection of expectation values and moments with
respect to a set of basic operators, $\hat{Q}$ and $\hat{P}$ in our case. We
therefore describe a (possibly mixed) state by an infinite set of numbers,
\begin{equation}
  Q=\langle\hat{Q}\rangle \quad,\quad P=\langle\hat{P}\rangle
\end{equation}
and central moments
\begin{equation}
  \Delta(Q^aP^b)= \langle(\hat{Q}-\langle\hat{Q}\rangle)^a
  (\hat{P}-\langle\hat{P}\rangle)^b\rangle_{\rm symm}
\end{equation}
in completely symmetric (or Weyl) ordering, for any positive integers $a$ and
$b$ such that $a+b>1$.

Expressed in these variables, our effective Hamiltonian reads
\begin{equation} \label{Heff}
  H_{\rm eff}(Q,P,\Delta(QP))= QP+\Delta(QP)\,.
\end{equation}
We need a Poisson bracket in order to generate Hamilton's equations. The
definition \cite{EffAc,Karpacz}
\begin{equation}
  \{\langle\hat{A}\rangle,\langle\hat{B}\rangle\}=
  \frac{\langle[\hat{A},\hat{B}]\rangle}{i\hbar}\,,
\end{equation}
combined with the Leibniz rule to apply it to products of expectation values
as they appear in moments, fulfills all the requirements for a Poisson
bracket. Moreover, it is easy to see that the usual evolution equations of
quantum mechanics (in Schr\"odinger or Heisenberg form) are equivalent to
Hamilton's equations generated by $H_{\rm eff}=\langle\hat{H}\rangle$ for
basic expectation values and moments.

Following this procedure, the present example of (\ref{Heff}) yields the
equations of motion
\begin{equation}
  \dot{Q}=Q \quad,\quad \dot{P}=-P
\end{equation}
without quantum corrections, accompanied by equations of motion
\begin{eqnarray*}
  \frac{{\rm d}\Delta(Q^2)}{{\rm d}\lambda} &=& 2\Delta(Q^2)\\
  \frac{{\rm d}\Delta(QP)}{{\rm d}\lambda} &=& 0\\
  \frac{{\rm d}\Delta(P^2)}{{\rm d}\lambda}&=& -2\Delta(P^2)
\end{eqnarray*}
as well as equations for higher-order moments. These equations can easily be
solved, determining how a state spreads out as $Q$ changes.

Unlike basic expectation values, which preserve the classical Poisson bracket
$\{Q,P\}=1$, moments are not canonical. (Their Poisson brackets are in general
quadratic in moments; see \cite{Karpacz,HigherMoments}.) The moment term in
(\ref{Heff}) therefore does not provide an intuitive picture as given by the
usual form of potentials in a Hamiltonian with a standard kinetic energy. In
this situation, it is convenient to transform moments to new canonical
coordinates, whose local existence is guaranteed by the Casimir--Darboux
theorem \cite{Arnold,Weinstein}. While it may in general be hard to derive
such coordinates, for second-order moments they have been found several times
independently
\cite{VariationalEffAc,GaussianDyn,EnvQuantumChaos,QHDTunneling,CQC,CQCFieldsHom}
(without drawing a connection with Poisson geometry): A canonical pair
$(s,p_s)$ such that $\{s,p_s\}=1$ together with the conserved quantity $U$ (a
Casimir function that has vanishing Poisson brackets with any other function)
produce the correct brackets of second-order moments if the latter are defined
as
\begin{eqnarray} \label{sps}
  \Delta(Q^2) &=& s^2\\
  \Delta(QP) &=& sp_s\\
  \Delta(P^2) &=& p_s^2+ \frac{U}{s^2}\,.
\end{eqnarray}
The uncertainty relation implies that $U$ is bounded from below by
$\hbar^2/4$.

In these variables, our effective Hamiltonian is given by
\begin{equation}
  H_{\rm eff}(Q,P,s,p_s)=QP+sp_s\,.
\end{equation}
It generates equations of motion that can be used to construct an extended
version of the analog harmonic oscillator if we simply copy the previous
procedure for the new canonical pair:
\begin{equation} \label{Hanalogeff}
  H_{\rm analog,eff}=\frac{1}{2}\left(\pi^2-Q^2+\pi_s^2-s^2\right)\,.
\end{equation}
Also the constraint $\pi=Q$ has to be doubled, accompanied by
$\pi_s=s$. Therefore, the two energy contributions from $(\pi,Q)$ and
$(s,\pi_s)$ have to vanish independently.

Notice that the effective analog Hamiltonian (\ref{Hanalogeff}) differs from a
standard effective formulation of the original analog Hamiltonian
(\ref{Hanalog}) in that it does not contain a term of the form $U/s^2$, as it
would be expected if $\pi^2$ were treated as an expectation value
$\langle\hat{\pi}\rangle^2+\Delta(\pi^2)$ of the kinetic energy, followed by
an application of the mapping (\ref{sps}) with $\pi$ instead of $P$. The
reason for this difference is the fact that $\pi$ is not an independent
momentum, owing to the constraint $\pi=Q$.

The analog Hamiltonian of $H_{\delta}$, however, is not subject to a sharp
constraint, such that the momentum $\pi$  remains independent. We can
therefore insert a $\pi$-version of (\ref{sps}) in the analog Hamiltonian,
producing
\begin{equation}
  H_{\delta,{\rm analog,eff}}= \frac{1}{2}\left(\pi^2+p_s^2+\frac{U}{s^2}-Q^2-s^2\right)\,.
\end{equation}

For small $Q$, the quadratic potential is replaced by a cubic one that
contains an absolute value. This potential is therefore not differentiable at
$Q=0$, preventing us from using an expansion
\begin{equation}
  \langle\hat{Q}^3\rangle=\langle (Q+\widehat{\Delta Q})^3\rangle= Q^3+
  3Q\Delta(Q^2)+\Delta(Q^3)
\end{equation}
of a differential cubic term, where $\widehat{\Delta Q}=\hat{Q}-Q$ such that
$\langle\widehat{\Delta Q}\rangle=0$. An extension to non-differential
potentials $V(Q)$ has been given by \cite{EffPotRealize} in the form
\begin{equation} \label{allorder}
  V_{\rm all-order}(Q,s)=\frac{1}{2}\left(\frac{U}{s^2}+V(Q+s)+V(Q-s)\right)
\end{equation}
based on a moment-closure condition that assumes a specific dependence of
higher-order moments on a single quantum variable, $s$. A similar potential
can also be derived from the Wigner function if additional conditions of
semiclassical behavior are imposed \cite{WignerSemiclass}.

\begin{figure}
  \begin{center}
    \includegraphics[width=16cm]{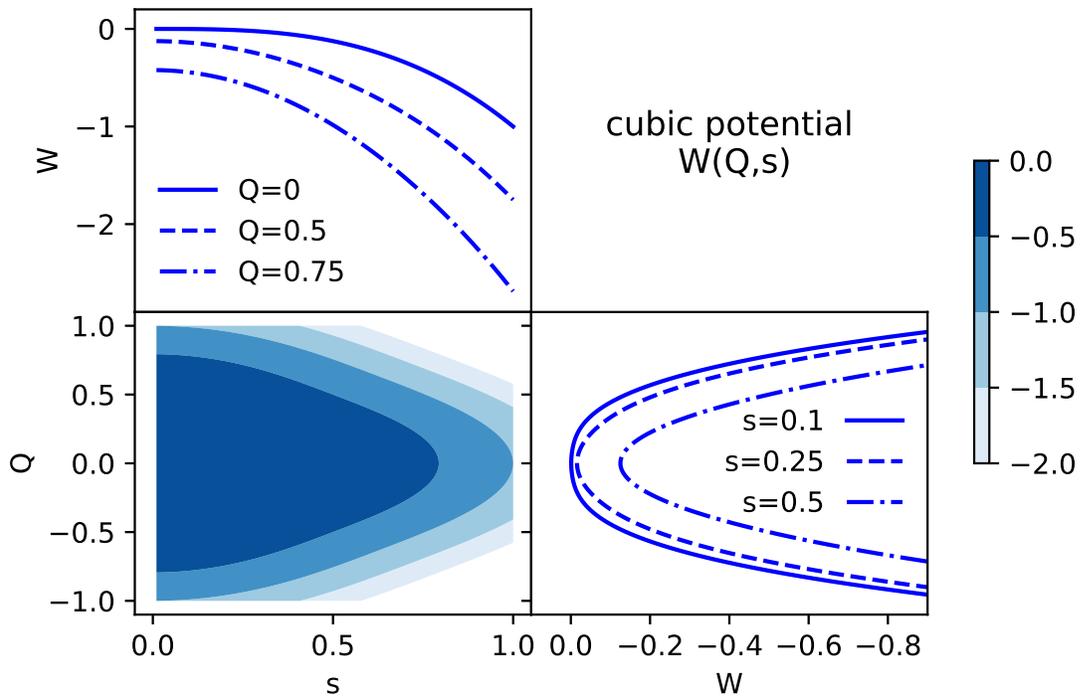}
    \caption{Visualization of the all-orders cubic potential with $\kappa=1$. A
      contour plot of the potential $W(Q,s)$ is shown at the bottom left,
      while cross-sections at constant $Q$ (top left) and constant $s$
      (bottom right), respectively, are appended along the $s$ and $Q$ axes. \label{f:Cubic}}
  \end{center}
\end{figure}

In cases in which $V(Q)$ is smooth,
a comparison of a Taylor expansion of the $V$-terms in (\ref{allorder}) with the
Taylor expansion of a generic effective potential,
\begin{eqnarray} \label{Veff}
  V_{\rm eff}(Q,\Delta(Q^n)) &=& \langle V(\hat{Q})\rangle= \langle
                                 V(Q+\widehat{\Delta Q})\rangle\nonumber\\
  &=& V(Q)+\sum_{n=2}^{\infty} \frac{1}{n!} \frac{\partial^nV(Q)}{\partial
      Q^n} \Delta(Q^n)\,,
\end{eqnarray}
shows that (\ref{allorder}) is implied by (\ref{Veff}) if the moments are such
that $\Delta(Q^n)=0$ for odd $n$, while $\Delta(Q^n)=s^n$ for even
$n$. Odd-order moments therefore do not contribute to (\ref{allorder}) for
smooth potentials, but they do in our case of a non-differentiable potential
because $\frac{1}{2}\left(|Q+s|^3+|Q-s|^3\right)$ contains cubic contributions
from $s$, which can easily be seen by evaluating the expression at $Q=0$. The
full dependence of the polynomial effective potential
\begin{equation} \label{W}
  W(Q,s)=\frac{1}{2}\left(V(Q+s)+V(Q-s)\right)
\end{equation}
on $Q$ and $s$ is illustrated in Fig.~\ref{f:Cubic} for the cubic case.

We relax the moment closure by introducing a free parameter, $a$, such that
$\Delta(Q^3)=a s^3$ and writing the effective Hamiltonian as
\begin{equation}
  H_{\rm cubic,eff}= \frac{1}{2}\pi^2+\frac{1}{2}p_s^2+\frac{U}{2s^2}- \kappa
  |Q|^3 - 3\kappa Q s^2- \kappa a s^3\,.
\end{equation}
The new parameter gives us more freedom to test the influence of the specific
state on the tunneling behavior.

\begin{figure}
  \begin{center}
    \includegraphics[width=16cm]{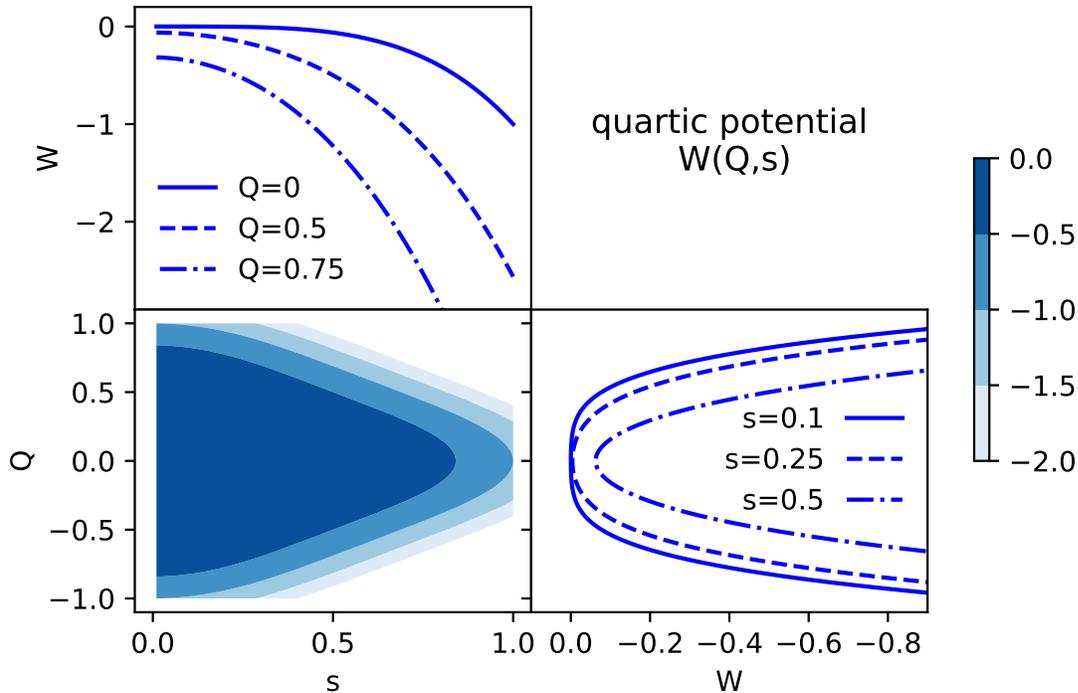}
    \caption{Visualization of the all-orders quartic potential with
      $\kappa=1$, using the method described in Fig.~\ref{f:Cubic}. \label{f:Quartic}}
  \end{center}
\end{figure}

\begin{figure}
  \begin{center}
    \includegraphics[width=16cm]{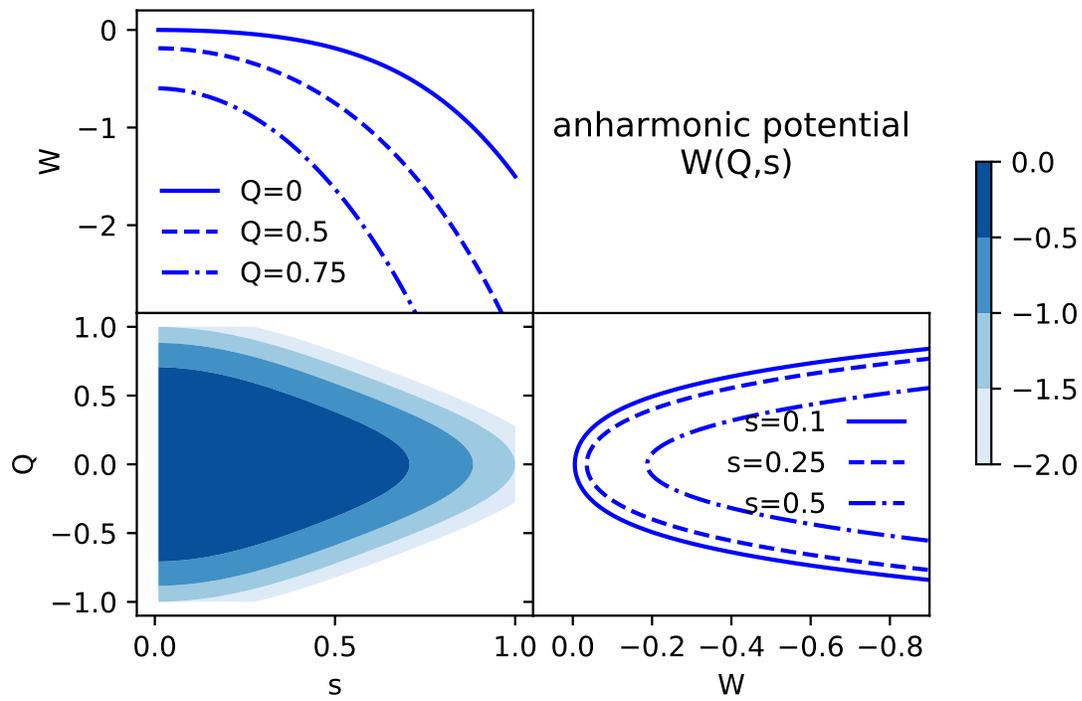}
    \caption{Visualization of an all-orders anharmonic potential with
      $\omega=\kappa=1$, using the method described in Fig.~\ref{f:Cubic}. \label{f:HarmonicQuartic}} 
  \end{center}
\end{figure}

A quartic or anharmonic potential, which may be relevant if quantization
ambiguities are taken into account, can be handled more easily because it is
smooth. The corresponding effective Hamiltonians are
\begin{equation}
  H_{\rm quartic,eff} = \frac{1}{2}\pi^2+\frac{1}{2}p_s^2+
  \frac{U}{2s^2}-\kappa Q^4- 6\kappa Q^2s^2- 4\kappa a
  Qs^3-\kappa b s^4
\end{equation}
and
\begin{equation}
  H_{\rm anharmonic,eff} = \frac{1}{2}\pi^2+\frac{1}{2}p_s^2+
  \frac{U}{2s^2}-\frac{1}{2}\omega^2(Q^2+s^2)-\kappa Q^4- 6\kappa Q^2s^2- 4\kappa a
  Qs^3-\kappa b s^4\,,
\end{equation}
respectively, where $a$ controls the cubic moment, as before, while $b$ plays
a similar role for the quartic moment assumed to be of the form
$\Delta(Q^4)=bs^4$. The polynomial parts of these two potentials, given in the
case of $a=0$ and $b=1$ by (\ref{W}) applied to a quartic and anharmonic $V$,
respectively, are shown in Figs.~\ref{f:Quartic} and \ref{f:HarmonicQuartic}.

\section{Tunneling dynamics}

Our effective potentials allow us to study various properties of evolving
quantum states while they are tunneling. As suggested by (\ref{Hvalue}), the
conserved energies in the quasiclassical models described by our effective analog
Hamiltonians are negative, while their potentials vanish at $Q=0$. In each
case, we therefore have a classical barrier that prevents $Q$ from changing
sign as it evolves. A reflection of $Q$ at its turning point corresponds to a
bounce in the quantum cosmological model represented by the analog
system. Quantum mechanically, however, tunneling is possible and may replace a
classical-type bounce with a transition through the singularity at $Q=0$.

The variable $Q$ in a quasiclassical formulation represents the expectation
value of an operator $\hat{Q}$ in an evolving state. A zero value of $Q$ does
not necessarily imply that the state is given by a wave packet centered around
the singularity. More likely in a tunneling scenario, the expectation value
might vanish because the wave function has split into two separate packets,
one that stays on the side of the barrier where the initial wave function was
set up, and one that has tunneled through the barrier. A vanishing expectation
value $Q$ therefore may not lead to as strong a singularity as in the
classical treatment. In this way, tunneling through the barrier could be
similar to a strict bounce behavior in which the wave function always stays on
one side of the barrier. Nevertheless, it is important to understand possible
tunneling dynamics as a first step in an analysis of the transition,
eventually with an inclusion of anisotropy and inhomogeneity. Some of the
relevant degrees of freedom may well be affected by tunneling through $Q=0$
even if the zero may refer only to an expectation value and not necessarily to
the support of a wave function. This possibility so far has not been
considered in the literature.

While an inclusion of anisotropy and inhomogeneity would be well beyond the
scope of the present paper, we proceed with a specific analysis of some
dynamical tunneling properties in our isotropic analog models. In particular,
we will focus on the question of how long it takes for a state to tunnel
through the classical barrier. The usual treatment of tunneling in quantum
mechanics, based on stationary states, does not reveal much about the actual
process. In a quasiclassical model, however, tunneling time is well defined
because we are able to track the evolving $Q$ and to check under which
conditions and for how long it moves through the range determined by the
classical barrier for a given energy. We will refer to this tunneling time as
$\Delta t_{\rm evolution}$.  As shown by
Figs.~\ref{f:Cubic}--\ref{f:HarmonicQuartic}, the effective potentials
decrease in the $s$-direction around local maxima of the classical
potential. Therefore, it is possible for certain trajectories to move around
the classical barrier in the $(Q,s)$-plane while maintaining the initial
energy.

Just as there is usually a non-zero reflection coefficient in the standard
treatment of tunneling, some of the quasiclassical trajectories do not move
around the barrier. We are interested in cases in which tunneling happens
predominantly, which requires a specific choice of initial values for $Q$, $s$
and their momenta. In order to ensure that a given trajectory tunnels, we pose
initial values at $Q=0$, at the maximum of the classical barrier. A given
negative energy value then requires a minimum value of $s$ for real momenta to
be possible. We choose an allowed value for $s$, set $p_s=0$ at the initial
time and solve the analog energy equation for $\pi$. A complete set of initial
values is thus obtained which can be integrated numerically, using equations
of motion generated by the corresponding effective Hamiltonian. The time
$\Delta t_{\rm evolution}$ required to cross the barrier can be read off from
the resulting trajectory.

Before we discuss our specific results, we point out that the time parameter
$t$ that appears in our tunneling times refers to evolution generated through
Hamilton's equations by an analog Hamiltonian. In general, this time is not
the proper time of the cosmological model described in analog form. Since we
are interested here in qualitative features of tunneling times, as in the
example that follow, a transformation from $t$ to proper time is not required.

\begin{figure}
  \begin{center}
    \includegraphics[width=16cm]{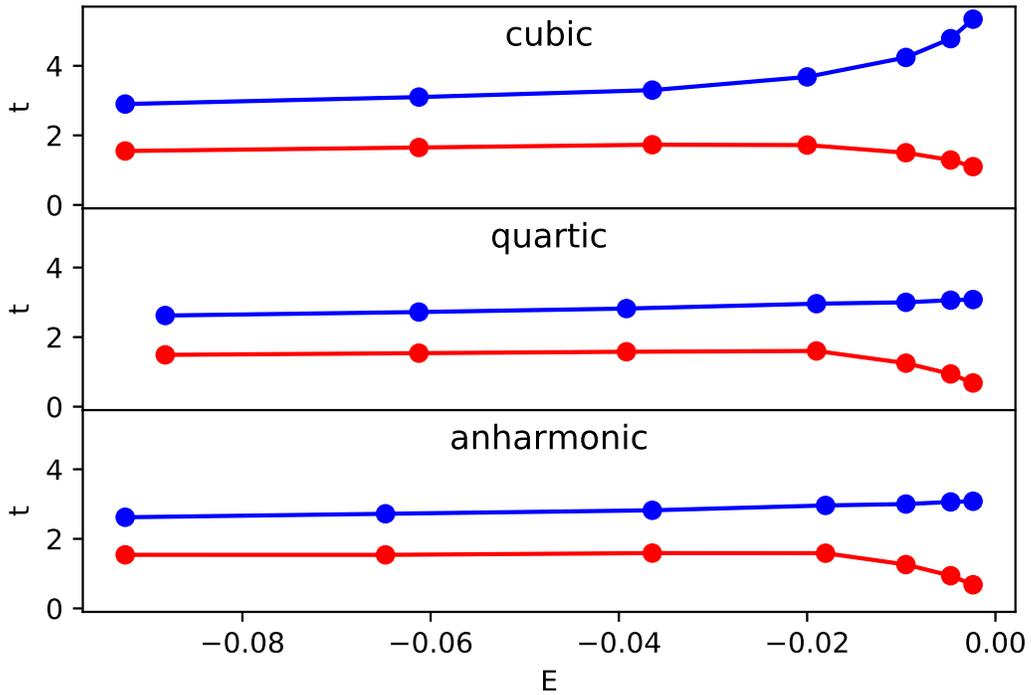}
    \caption{Tunneling times for various potentials, using two different
      methods. The red curves show our results obtained from evolution in a
      quasiclassical model. The blue curves show an alternative definition
      (\ref{tintegration}) based on integrating ${\rm d}t=(m/|p(Q)|){\rm d}Q$
      across the classical barrier, using the imaginary momentum $p(Q)$. While
      both definitions lead to comparable orders of magnitude, their trends in
      particular at small energies are markedly different. The quasiclassical
      tunneling time does not significantly depend on the specific
      potential. \label{f:CubicQuartic}}
  \end{center}
\end{figure}

\begin{figure}
  \begin{center}
    \includegraphics[width=14cm]{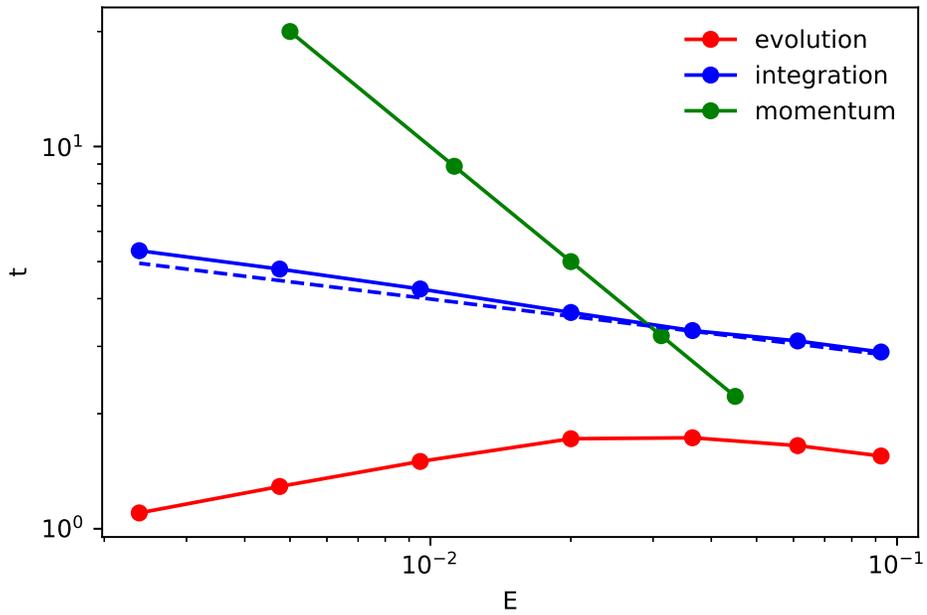}
    \caption{Tunneling times for the cubic potential in logarithmic display,
      showing results from quasiclassical evolution (red), integration of
      ${\rm d}t=(m/|p(Q)|){\rm d}Q$ (blue), and a simple estimate
      $m\Delta Q/p_0$ based on the momentum $p_0$ at the maximum of the
      barrier of width $\Delta Q$ (green). The blue dashed line is not a fit
      but has been included for orientation, showing the curve
      $t(E)=2E^{-0.15}$ for comparison. \label{f:CubicLog}}
  \end{center}
\end{figure}

Our results are shown in Figs.~\ref{f:CubicQuartic} and
\ref{f:CubicLog}. Figure~\ref{f:CubicQuartic} demonstrates that the dependence
of tunneling times on the energy does not significantly depend on the specific
low-order polynomial potential. Interestingly, this feature is not shared by
alternative definitions of tunneling times, in particular at small absolute
values of the energy, close to the top of the barrier. The figure shows one
alternative method for comparison, given by a more traditional proposal
\cite{TunnelingReview,BarrierTime,AttoclockReview} to estimate the tunneling
time by the integral
\begin{equation} \label{tintegration}
  \Delta t_{\rm integration}= \frac{1}{\sqrt{2}}\int_{Q_-}^{Q_+} \frac{\sqrt{m}
    {\rm d}Q}{\sqrt{V(Q)-E}}
\end{equation}
where $Q_{\pm}$ are the classical turning points for the potential $V(Q)$ at
energy $E$. This definition assumes that the imaginary part of the solution
for $p$ of the energy equation $E=(2m)^{-1}p^2+V(Q)$ under the barrier may
provide a reasonable estimate for $m{\rm d}Q/{\rm d}t$.

As our figure shows, the order of magnitute is close to our more detailed
tunneling time based on quasiclassical evolution, but it is rather different
at small energies, close to the top of the barrier of our potential. The
evolution time clearly approaches zero as the energy gets closer to the top of
the barrier because the required distance to travel eventually vanishes and
momenta stay finite. But while the integration range in (\ref{tintegration})
also shrinks as $Q_-\to Q_+$ when $E$ approaches zero, the integrand gets bigger
because it diverges at the turning points. This trend is particularly clear
for the cubic potential in Fig.~\ref{f:CubicQuartic}.

The logarithmic plot of Fig.~\ref{f:CubicLog}, shown only for the cubic
potential, confirms this behavior. While the integration time, by inspection, is close to a
power law $t(E)\sim 2E^{-0.15}$, the evolution time is not of power-law form
and exhibits a more complicated behavior. In addition to these two definitions,
the figure also shows a rough estimate of the tunneling time,
\begin{equation}\label{tmomentum}
  \Delta t_{\rm momentum}= \frac{m\Delta Q}{p_0}
\end{equation}
based on the width $\Delta Q=Q_+-Q_-$ of the barrier and the momentum $p_0$ at
$Q=0$ obtained from a given energy in a quasiclassical model. (The value of
$p_0$ would be used as an initial condition in the more precise definition of
tunneling times by complete evolution.) This behavior is of power-law form and
does not agree with any one of the other two definitions. The precise
evolution of a quasiclassical model is therefore relevant, and not just the
fact that the energy can be above the potential for sufficiently large
$s$. The latter feature, which guarantees that a real $p_0$ is obtained at
$Q=0$, is shared by the evolution time and the momentum time, and yet the two
results are different in many respects.

\section{Discussion and implications for quantum cosmology}

We have analyzed a detailed definition of tunneling time in analog models of
quantum cosmological bounces. The analog method allowed us to model dynamics
initially given for variables of an ${\rm sl}(2,{\mathbb R})$ algebra in terms
of standard Hamiltonians. The Hamiltonians being of standard form, given by a
kinetic energy quadratic in the momentum and a low-order polynomial potential,
it becomes possible to construct a direct application of methods developed for
quantum mechanics. In a second step, we applied quasiclassical formulations of
non-adiabatic quantum dynamics in order to reveal crucial dynamical features
of quantum evolution.

Our analysis revealed several properties of relevance for quantum
cosmology. In particular, we showed that the definition of tunneling time
based on non-adiabatic evolution of expectation values and moments of a state,
unlike other and more traditional definitions, leads to small tunneling times
near the top of the barrier. Moreover, as the energy decreases and the barrier
becomes larger, the tunneling time plateaus or even starts decreasing as seen
easily in the logarithmic plot of Fig.~\ref{f:CubicLog}. Even a large barrier,
therefore, does not present an effective way to stabilize bounce models
dynamically. A classical barrier around the singularity does not suffice to
prevent a dynamical transition through the singularity.

Whether this transition can re-introduce singular effects for instance in the
dynamics of anisotropy or inhomogeneity remains an important open
question. The variable $Q$ in a quasiclassical model represents the
expectation value of the volume operator. The value zero may be reached for
various reasons in terms of an underlying wave function, for instance by a
wave function centered at zero or by a suitable superposition of two wave
packets centered at non-zero values even if their support does not include
zero. Which possibility is generically realized could be revealed by a
detailed analysis of the fluctuation variable, $s$, in quasiclassical
models. Our preliminary examples usually require a large value of $s$ at $Q=0$
because the effective potential has to drop down sufficiently far in the
$s$-direction to give way for the quasiclassical trajectory. However, the
required minimum of $s$ depends on the energy, which in turn depends on
detailed quantum parameters of the underlying cosmological model. Different
versions of tunneling wave functions may therefore be realized.

\section*{Acknowledgements}

This work was supported in part by NSF grant PHY-1912168.


\end{document}